# Unambiguous identification of the indirect band nature of atomically thin hexagonal boron nitride

Lei Fu[1†], Yuqing Hu[4†], Ning Tang[1,2,3*], Junxi Duan[4*], Xionghui Jia[1], Huaiyuan Yang[1], Zhuoxian Li[1], Xiangyan Han[1], Guoping Li[1], Jianming Lu[1], Lun Dai[1,2], Weikun Ge[1], Bo Shen[1,2,3*]

[1]State Key Lab for Mesoscopic Physics and Frontiers Science Center for Nano-optoelectronics, School of Physics, Peking University, Beijing, China.
[2]Collaborative Innovation Center of Quantum Matter, Beijing, China
[3]Peking University Yangtze Delta Institute of Optoelectronics, Nantong, Jiangsu, China
[4]Key Laboratory of Advanced Optoelectronic Quantum Architecture and Measurement (MOE), School of Physics, Beijing Institute of Technology, Beijing, China

†Lei Fu and Yuqing Hu contributed equally to this work.
*Corresponding author: ntang@pku.edu.cn, junxi.duan@bit.edu.cn, bshen@pku.edu.cn.

## Abstract

Atomically thin hexagonal boron nitride (h-BN), especially monolayer, has garnered increasing attention due to its intriguing optical and light-matter-interaction properties. However, its intrinsic optical properties and electronic band structure, have long remained elusive. In this study, near-resonance excited deep-UV photoluminescence/Raman spectroscopy and deep-UV reflectance contrast spectroscopy are utilized to experimentally investigate the optical properties of atomically thin h-BN across various layer numbers. It is revealed that the absence of luminescence in 1-3 layers h-BN is indicative of their indirect band gap nature, rectifying previously adopted identification of a direct band gap in monolayer BN. Notably, band-edge luminescence signals and indirect bandgap absorption start to appear in 4-layer, and the luminescence intensity increases with the number of layers, suggesting that interlayer interactions and periodicity along the z-axis enhance phonon-assisted indirect bandgap transition, even in the 4-layer case, and furthermore indicating the formation process of flat bands at the K and M valleys as the periodicity along the z direction increases. Additionally, the prominent resonance Raman signals in atomically thin h-BN underscore strong electron-phonon coupling in this material.

Hexagonal boron nitride (h-BN) has garnered considerable attention over the decades due to its unique properties, including its role as the cleanest dielectric for 2D materials and its ultra-wide bandgap with exceptional optical characteristics[1, 2]. Bulk h-BN exhibits remarkably high internal quantum efficiency (IQE)[3], comparable to or even exceeding that of direct-bandgap semiconductors, making it a promising candidate for deep-UV LEDs and lasers[4-6]. The bandgap of bulk h-BN is known to be indirect, with the conduction band minimum at M point and the valence band maximum at K point in the first Brillouin zone[7, 8], while the high IQE is attributed to the huge oscillator strength of phonon-assisted transitions originated from the flat bands along z direction at K and M valleys[9].

Beyond bulk materials, atomically thin h-BN, especially monolayer, has attracted significant interest in recent years due to its intriguing optical and light-matter-interaction properties, such as large Fröhlich polarons[10], highly confined phonon polaritons[11, 12], and ultrabright single-photon emission (SPE) from color centers[13, 14]. Twisted homostructures and heterostructures based on atomically thin h-BN are also in the spotlight, such as rhombohedral boron nitride (rBN) with outstanding optical nonlinearity[15], twisted h-BN/graphene heterostructures with tunable interband transitions[16], and twisted h-BN with enhanced colour center emission at the interface[17]. Particularly, monolayer BN is predicted to be the most promising candidate for the strong co-occurrence of bright s-like and bright p-like states in an intrinsic 2D material[18], and holds potential for lightwave valleytronics in insulating quantum systems[19].

Although many of these intriguing properties of the very few-layer h-BN have been explored, the very fundamental understanding of its bandgap nature and intrinsic optical properties, especially those of the monolayer BN, have long remained elusive, with conflicting reports leading to serious ambiguity. Theoretical calculations have predicted a direct bandgap at K valley for monolayer BN and an indirect bandgap similar to bulk h-BN for layer number above 2[20-22]. In terms of the experimental approach, light emission from h-BN thicker than 6 layers has been observed using cathodoluminescence (CL) spectroscopy[23]; notably the monolayer BN epitaxially grown on graphite substrate has been shown to have a direct bandgap at K valley, with the exciton emission energy around 6.1 eV[24-29]. As for monolayer BN on an insulating substrate, which reflects its intrinsic properties, the only reported photoluminescence (PL) signals have also suggested a direct bandgap[30]. However, due to the theoretically predicted huge exciton binding energy of ~2 eV[18, 21], the excitation mechanism is doubtful, and the fine structure of the band-edge luminescence, with many enigmatic satellites, is complex and hard to interpret[30, 31]. On the other hand, further theoretical work using the GW approximation suggests a different scenario: monolayer BN possesses an indirect bandgap with the conduction band minimum at Γ point while the valence band maximum at K point[18, 32-36], and the interaction with graphite substrate can lower the direct gap at K point much more than the indirect one and thus tune it to a direct bandgap at K point[35]. This discrepancy necessitates a clear and definitive identification of its band nature, which is crucial for both fundamental understanding and guiding future physical research and applications of monolayer BN.

In this article, the optical properties of h-BN on sapphire substrates with various layer numbers (1~6 and 12 layers) have been investigated using deep-UV photoluminescence/Raman spectroscopy under near-resonance excitation, along with deep-UV reflectance contrast spectroscopy. As a matter of fact, for h-BN thinner than 3 layers, only a series of resonance Raman signals are observed, proving their indirect bandgap nature. As the layer number approaches 4 and more, the band-edge luminescence emerges while the resonance Raman signals become weaker with increasing layer numbers. Moreover, phonon-assisted absorption by indirect bandgap emerges for 4 layers and thicker, consistent with the PL result, indicating that interlayer interactions and periodicity along z-axis enhance the phonon-assisted indirect bandgap transition even in 4-layer. Other than the band structure, our resonance Raman results also reveal the main high-order Raman modes, providing an excellent platform for studying the electron-phonon coupling in h-BN[37].

**Sample characterization**

Atomically thin BN across various layer numbers is mechanical exfoliated onto a Si/SiO$_2$ substrate with 90 nm thick SiO$_2$. The layer numbers are determined by combination of Raman and Atomic Force Microscope (AFM) test. Due to the substrate effect and thin surface contamination, the layer thickness obtained by AFM is usually a little larger[38], so it cannot give the exact layer number of atomically thin BN. On the other hand, the intensity of the Raman peak around 1366 cm$^{-1}$ (G band) of few-layer h-BN has been proved to have a linear dependence on the layer number[38, 39], indicating that Raman is a powerful technique for layer number identification. We obtained a stepped sample with five different layer numbers (Fig. 1a). The steps between different layer numbers are characterized by AFM (Fig. 1a), along with the intensity of the G band at the five different regions behaving 1:2:3:4:5 (Fig. 1b), indicating that the five regions correspond to 1~5 layers (1~5L). Using this sample as a reference, we prepared additional seven isolated BN flakes without step for research in the main text. The layer numbers of the seven BN flakes are identified to be 1~6L and 12L, referring to the intensity of the G band of the reference sample (inset of Fig. 1b).

The 1~6L and 12L BN flakes are then transferred to a sapphire substrate using polycarbonate (PC) as an intermedium (see the section "Methods"). There are two advantages for using sapphire as substrate, rather than the original Si/SiO$_2$ substrate. One is the convenience of absorption spectrum, as the reflectance contrast is directly proportional to the absorption for thin film on transparent substrates[40-42]. Another advantage is better protection of samples[43], as the thermal conductivity of sapphire is far better than SiO$_2$ at low temperature[44, 45], and no heating of sapphire under UV radiation due to its high energy band gap[43].

### Direct K-K and phonon-assisted indirect M-K absorption

As mentioned above, the reflectance contrast spectrum on transparent substrates can be taken as an absorption spectrum. The normalized reflectance contrast spectra of 1~6L and 12L BN on sapphire substrate are shown in Fig. 2a. The absorption curves are deconvoluted into four peaks at most, marked as A1, A2, A3, A4, respectively. For 1~3L, only A1 and A2 exist. With increasing layer number, A3 emerges in 4L, and A4 appears only in the 12L BN. The energies of the four peaks are presented in Fig. 2b, which enable us to make a clear assignment, except for A1, as it is too broad to make the fitting energy exactly. The physical processes of light emission and absorption in h-BN are illustrated in Fig.2c and 2d, respectively.

First, the A2 peak around 6.14 eV can be assigned to the direct K-K absorption in h-BN for all layer numbers. The energy of the A2 peak in monolayer BN (6.15 eV) is consisted with previous reports[46]. The A3 peak, with energy around 6.109eV, can be assigned to phonon-assisted indirect M-K absorption. For optical response of an indirect bandgap, there exists a mirror symmetry between the absorption and emission energies, with the former blue shifted by a phonon energy and the latter red shifted by the same phonon energy, referring to the indirect bandgap energy[8]. Considering the indirect exciton energy of 5.955eV for bulk h-BN[8], this absorption energy around 6.109eV is symmetrical with the emission energy of 5.801eV in 12L BN (Fig. 3n), thus confirming it to be an indirect M-K absorption assisted by TO/LO phonons around T points[8] in the first Brillouin zone (inset of Fig. 2b). Similarly, the A4 peak with energy around 6.075eV, is estimated to be an indirect M-K absorption assisted by TA/LA phonons around T points[8] in the first Brillouin zone. However, the calculated emission energy using the mirror symmetry 2×5.955-6.075=5.835 eV, is significantly deviated from the measured PL energies of 5.879 eV for $X_{LA(T)}$ (see Fig.2c: exciton recombination assisted by LA(T) phonon, similar below) and 5.905 eV for $X_{TA(T)}$. The main reason for this deviation might be due to the uncertainty of the fitting for broad peaks. While from another perspective, the appearance of A3 and A4 is consistent with the PL measurement in the next section, suggesting that the interlayer interactions and periodicity along z-axis enhance the phonon-assisted indirect bandgap transitions even in 4-layers.

As for the A1 peak, with the largest FWHM and uncertainty during the fitting process, we can only make a suggested identification. The most probably origin of this peak might be a phonon-assisted direct K-K absorption, with a phonon emitted during the process. As the energy difference between A1 and A2 peak is about 125 meV, corresponding to the Raman shift of about 1008 cm$^{-1}$, it is close to 942.2 cm$^{-1}$ (ZO(M)+ZA(M)) or 1066.6 cm$^{-1}$ (2TA(M)) peaks in the resonance Raman spectrum of the monolayer BN (see the section "Preliminary assignments of the resonance Raman peaks"). Considering the extremely weak intensity of the 1066.6 cm$^{-1}$ peak, the involved phonons are more probably to be ZO(M)+ZA(M). Although there also exists a significantly deviation, considering the extraordinary FWHM of this peak and the resulted huge uncertainty of fitting, the deviation seems acceptable.

### Identification of the indirect band nature

The PL and resonance Raman signals of 1~6L and 12L h-BN flakes are acquired using our homemade deep-UV photoluminescence spectroscopy system, with the excitation wavelengths around 197nm (6.294 eV) for near-resonance excitation, just on the higher energy shoulders of the absorption curves in Fig. 2a. The excitation power is about 30μW to protect the samples from being damaged by laser. All the tests are conducting under 4K temperature for best signal-to-noise ratio (for more information, see the section "Methods").

For each sample, at least 2 excitation wavelengths are applied to distinguish the PL and resonance Raman signals. The results are illustrated in Fig. 3a-n, with backgrounds subtracted (Supplementary Note 1). For 1~3L h-BN, the spectrum under two different excitation energies shifted globally by 13 meV (just the energy difference of excitation laser) in the energy coordinates, but consistent in the Raman shift coordinates, confirming that only resonance Raman signals exists for 1~3L h-BN (neglecting some faint luminescence signals below 5.75 eV, which is probably originated from defects). While for 4~6L h-BN, the spectrum under different excitation energies can neither be fully consistent in the energy coordinates nor in Raman shift coordinates (For normalized comparison, see Supplementary Note 2). To see clearly, four excitation wavelengths are used for 4~6L h-BN. It can be seen that the components of PL increase with increasing layer numbers, and become obvious for 6L. For 12L h-BN, the spectrum becomes dominated by PL, mainly consisting of phonon-assisted indirect emission ($X_{LA(T)}$/$X_{TA(T)}$/$X_{LO(T)}$/$X_{TO(T)}$) and a series of defect induced emissions at lower energies, the same as that of bulk h-BN[8], except for the appearance of surface exciton ($X_S$), which have been reported to exist in few-layer h-BN[3], and make the $X_{TA(T)}$ hard to be separated clearly. Meanwhile, the intensity of the resonance Raman signals becomes negligible for 12L h-BN. The quench of the resonance Raman signals may be attributed to the reduced exciton lifetime due to increased radiative recombination rate with flat bands forming along z direction at K and M valleys. The exclusive relationship is discussed in detail as follows.

To make a better clarification, it is necessary to figure out the relationship between the intensity of resonance Raman/PL and layer number. The PL together with resonance Raman signals of 1~6L and 12L h-BN under excitation energy of about 6.300 eV are illustrated in Fig. 3o. It can be seen that the components of the resonance Raman signals between 1100 and 1700 cm$^{-1}$ are separated far from the PL signals. So the Raman signals of different layer numbers are integrated over 1100~1700 cm$^{-1}$, and the result is illustrated by the red points in Fig. 4. For extraction of the PL components, things become complex because the PL signals mix up with the resonance Raman signals at each excitation energy. However, the shape of the resonance Raman spectrum seems consistent for all layer numbers, except for a few sharp modes (Supplementary Note 3). So we tried to subtract the spectrum of the 3L sample from the spectrum of the 4~6L samples, followed by a fitting, to get an approximate PL spectrum of 4~6L (Supplementary Note 4). The approximate PL spectrum, together with the whole spectra (including PL and resonance Raman) of 12L h-BN, are illustrated in Fig. 3p. While the integrated intensities over 5.7~6.0 eV of different layer numbers are illustrated by the blue points in Fig. 4. With increasing layer number, the intensity of the PL increases, whereas the intensity of the resonance Raman signals decreases. The Raman scattering probability in a resonance Raman process can be expressed by the following formula[47]:

$$P \approx \left(\frac{2\pi}{\hbar}\right) \left| \frac{\langle 0|H_{eR}(\omega_s)|a\rangle\langle a|H_{e-ph}|a\rangle\langle a|H_{eR}(\omega_i)|0\rangle}{(E_a - \hbar\omega_i - i\Gamma_a)(E_a - \hbar\omega_s - i\Gamma_a)} \right|^2 \quad (1)$$

Here $H_{eR}$ denotes the electron-radiation interaction Hamiltonian, $H_{e-ph}$ denotes the electron-phonon interaction Hamiltonian, $|0\rangle$ is the initial and final state, $|a\rangle$ is an intermediate electronic state and $E_a$ is the corresponding energy level, $\hbar\omega_i$ is the energy of the incident photons, $\hbar\omega_s$ is the energy of the scattered photons, and $\Gamma_a$ is the damping constant, related to $\tau_a$ by $\Gamma_a = \hbar/\tau_a$, where $\tau_a$ is the lifetime of the intermediate electronic state $|a\rangle$.

With increasing layer numbers, the increasing PL intensity indicates a shorter lifetime of the photogenerated excitons, and thus a larger $\Gamma_a$ in the denominator of formula (1). That explains why the intensity of the resonance Raman signals decreases with increasing layer numbers and disappears in 12L or bulk h-BN.

As mentioned before, the strong luminescence in bulk h-BN has been attributed to the giant density of states (DOS) arising from the flat bands at K and M valleys along z-axis, which enhance the phonon-assisted indirect bandgap transition. However, for structures with limited periodicity, such as 4 to 12 layers, to claim the conception of flat bands is not precise due to the shorter periodicity along the z-axis. Nonetheless, it is important to note that real crystals all have finite periodicity, and the formation of flat bands, i.e. increasing of DOS, is a gradual process rather than an abrupt transition at a specific layer number. In this study, the observation of both phonon-assisted indirect bandgap luminescence and absorption in the 4-layer h-BN indicate that the interlayer interactions and periodicity along the z-axis indeed play a role in increasing the DOS, and making it deviate from the typical nature of an indirect bandgap semiconductor. As for 12 layers, which corresponds to just six periods along the z-axis, this effect has already become significant, enhancing the luminescence spectrum closely resemble that of bulk h-BN.

Thus, although the flat bands might have not formed in these few-layer h-BN structures, the evolution of their photoluminescence (PL) spectra strongly suggests the gradual increasing DOS due to interlayer interactions and periodicity along the z-axis. In contrast, without this enhancement, the 1~3L h-BN behave as typical indirect bandgap semiconductors, with no band-edge luminescence. Therefore, we conclude that h-BN exhibits an indirect bandgap regardless of the number of layers. The key difference lies in the fact that, for 4 layers and above, interlayer interactions and periodicity along the z-axis increase the DOS and enhance phonon-assisted transitions, whereas this enhancement is absent in 1~3 layers, explaining the lack of band-edge luminescence under near-resonant excitation in these thinner h-BN samples.

## Preliminary assignments of the resonance Raman peaks

The resonance Raman spectrum of monolayer BN is deconvoluted into 20 peaks (the last one does not seem to be effective signal, and is only used to fit the curve), as shown in Fig. 5. Referring to the experimentally obtained phonon dispersion curve of monolayer BN[48], together with the calculated density of states (DOS) in monolayer BN[49], the source of the 19 resonance Raman peaks can be well assigned, as shown in Table 1.

TABLE 1. Assignments of experimentally observed resonance Raman peaks in monolayer BN. The reference frequences are taken from Ref.48.

| Index | Raman Shift ($cm^{-1}$) | Preferred Phonon modes | Reference ($cm^{-1}$) | Alternative Phonon Modes | Reference ($cm^{-1}$) |
|---|---|---|---|---|---|
| 1 | 650.7 | 2ZA(MK) | 599.6 | ZO(M) | 620.6 |
| 2 | 809.0 | TA(M)+ZA(M) | 842.1 | ZO(Γ) | 797.0 |
| 3 | 942.2 | ZO(M)+ZA(M) | 918.0 | | |
| 4 | 1066.6 | 2TA(M) | 1089.4 | LA(K) | 1059.1 |
| 5 | 1253.0 | 2ZO(M) | 1241.2 | TO(M) | 1252.8 |
| 6 | 1293.7 | LA(K)+ZA(K) | 1361.2 | LO(M) | 1297.8 |
| 7 | 1375.4 | LO/TO(Γ) | 1352.5 | LA(K)+ZA(K) | 1361.2 |
| 8 | 1537.2 | TO(MK)+ZA(MK) | 1560.6 | | |
| 9 | 1925.9 | LA(K)+TA(K) | 1923.4 | LO(M)+ZO(M) | 1918.4 |
| 10 | 2166.5 | TO(K)+TA(K) | 2133.1 | 2LA(K) | 2118.3 |
| 11 | 2308.6 | TO(K)+LA(K) | 2327.9 | 2LA(M) | 2302.9 |
| 12 | 2561.9 | 2TO(MK) | 2521.6 | | |
| 13 | 2772.2 | 2LO/TO(Γ) | 2705.0 | | |
| 14 | 3064.5 | 2TO(MK)+2ZA(MK) | 3121.2 | LO/TO(Γ)+LA(K)+ZO(K) | 2999.7 |
| | | | | LO/TO(Γ)+LA(M)+TA(M) | 3048.6 |
| 15 | 3226.4 | LO/TO(Γ)+TO(M)+ZO(M) | 3225.9 | LO/TO(Γ)+TO(K)+ZO(K) | 3209.4 |
| 16 | 3461.3 | LO/TO(Γ)+2LA(K) | 3470.8 | LO/TO(Γ)+TO(K)+TA(K) | 3485.6 |
| 17 | 3856.2 | LO/TO(Γ)+2TO(MK) | 3874.1 | 2LA(K)+2TA(K) | 3846.8 |
| 18 | 4040.8 | 3LO/TO(Γ) | 4057.5 | | |
| 19 | 4464.4 | 2LO(K)+2LA(K) | 4479.0 | | |

For high-order Raman signals, the identification is usually complex due to multiple combinations of phonons at various high symmetry points. We carefully examined all possible combinations of phonons, and give the most probable assignments due to the following principles. First, the phonon modes that keep conservation of momentum are preferred. Those modes with momentum compensated by defects[47] are placed in "Alternative Phonon Modes". Second, the mode ZO(Γ) is forbidden by Raman selection rules, however, it can be brightened by resonance with dark excitons[47], so we place it in "Alternative Phonon Modes" because the energy level of the dark excitons cannot be determined in this experiment. Third, those phonon modes with high DOS are preferred, and those with low DOS are placed in "Alternative Phonon Modes" too. Fourth, the energy of peak 14 is twice that of peak 8, and only the two peaks blue shifts significantly with increasing layer numbers (Fig. 3o). Thus it is quite reasonable to assign peak 14 to overtones of peak 8, i.e. 2TO(MK)+2ZA(MK).

The blueshifts of peak 8 and peak 14 are due to Davydov splitting, which is usually more sensitive to out-of-plane modes[50]. However, peak 8 and peak 14 are not the only two peaks that involve out-of-plane modes. The reason why Davydov splitting only exists for the two peaks may be originated from the flat phonon bands along M-K. Flat bands result in high DOS that contributes to Raman scattering, and thus more sensitive to interlayer coupling.

Overall, this is the first detailed resonance Raman spectrum for monolayer BN. Various high-order peaks indicate that strong electron-phonon interaction occurs in this system, again supporting the phonon-assisted excitation mechanism of monolayer BN. Due to the theoretically predicted huge exciton binding energy (about 2 eV)[21], the first excited state locates at about 0.9 eV above the 1s state of exciton (corresponding to about 177 nm), far beyond the reach of our fourth harmonic of Ti: sapphire oscillator. However, the absorption spectrum shows that there still exists a weak absorption till 6.4 eV, due to the phonon-assisted direct K-K absorption (broad A1 peak in Fig. 2a).

It is worth noting that, the assignments of the peaks are still a kind of preliminary, as more detailed information referring the exact electron-phonon coupling modes needs to be examined by polarization resolved and circular polarization resolved resonance Raman in future. Nevertheless, the fundamental recognition of the bandgap nature is reliable.

In summary, this work provides an unambiguous identification of the indirect bandgap nature of atomically thin h-BN by using near-resonance excited deep-UV photoluminescence/Raman spectroscopy combined with deep-UV reflectance contrast spectroscopy. Neither band-edge luminescence signals nor indirect bandgap absorption exists for 1~3L h-BN, similar to traditional indirect bandgap semiconductors. In contrast, PL signals and phonon-assisted indirect M-K absorption emerge in 4-layer and thicker h-BN, suggesting that interlayer interactions and periodicity along the z-axis enhance phonon-assisted indirect bandgap luminescence even in 4-layer h-BN. This also indicates the gradual formation of flat bands at the K and M valleys as the periodicity along the z-axis increases. Additionally, strong electron-phonon coupling in atomically thin h-BN has been preliminary demonstrated experimentally by using near-resonance Raman spectroscopy. This work provides crucial insights into the electronic band structures and optical properties of atomically thin h-BN, especially for monolayer, and thus paves the way towards further experimental research into the optoelectronics of monolayer BN, including deep-UV valleytronics and co-occurrence of bright s-like and bright p-like states. That is, bandgap engineering such as dielectric environment modulation and strain modulation for revealing its direct bandgap at K point.

# Methods

### Sample fabrication

Atomically thin h-BN across various layer numbers is mechanically exfoliated onto Si/SiO$_2$ substrates with 90 nm thick SiO$_2$. The raw h-BN crystal is commercial h-BN crystal from Onway Technology. The contrast of the microscope is adjusted for better visibility. The hBN flakes are picked up by a stamp consisting of polycarbonate (PC) on polydimethylsiloxane (PDMS), and then released onto pre-marked sapphire substrates. The samples are soaked in chloroform to remove PC. Once transferred to the sapphire substrate, these samples become totally invisible and can only be located by help of the marks.

### Atomic force microscopy

AFM measurements were performed in ambient conditions using an Asylum Research Cypher-S system. Tapping-mode is taken for confirming the step between different layer numbers.

### Raman spectroscopy in the visible range

The Raman spectroscopy in the visible range is taken at room temperature for identification of layer numbers of exfoliated h-BN. A continuous-wave laser with wavelength of 532 nm and stable power of 1 mW was focused on the sample by an 100x objective (50x; N.A. = 0.9), and the focus spot size is about 800 nm. The spectrum are collected by a spectrometer (iHR550, Horiba) equipped with a 1200 grooves/mm$^{-1}$ grating blazed at 550 nm and a liquid nitrogen cooling CCD (Symphony II BIUV, Horiba). The exposure time is set to 1 min for better signal-to-noise ratio.

### Deep-UV PL/Raman spectroscopy and deep-UV reflectance contrast spectrum

All optical experiments are conducted in a homemade deep-UV integrated optical system. The samples are mounted in a helium-free cryostat system (ST-500-P, Janis), and the temperature is fixed at 4K for all deep-UV optical experiments. In deep-UV PL/Raman spectroscopy, the excitation laser around 197 nm is provided by the fourth harmonic generation of a tunable Ti:sapphire laser with a repetition rate of 80 MHz and pulse duration ∼140 fs. The deep-UV laser is then focused on samples by a 36x reflective microscope objective (N.A. = 0.5), and the laser spot size is about 200nm. The spectrum are collected by a spectrometer (iHR550, Horiba) equipped with a 2400 grooves/mm$^{-1}$ grating blazed at 330 nm and a liquid nitrogen cooling CCD (Symphony II BIUV, Horiba). The laser power is set to 30μW to avoid damage to the sample. The exposure time is set to 5 min for better signal-to-noise ratio. A customized long-pass edge filter with a 50% transmission edge at 198.5 nm is used for deep-UV resonance Raman spectroscopy.

For deep-UV reflectance contrast spectrum, all remains the same except for taking a deuterium lamp as light source and the exposure time is set to 10s. The focused spot size of the deuterium lamp is about 10 μm, sometimes larger than the prepared h-BN plates, and the absolute intensity of reflectance contrast spectrum is not comparable because of different irradiated areas between different layer numbers, so all the reflectance contrast spectrum are normalized.

# Data availability

The data are available from the corresponding authors on reasonable request. Source data are provided with this paper.

## Acknowledgements

This work was supported by the National Key Research and Development Program of China (Nos. 2022YFB3605600, 2022YFA1604302, and 2020YFA0308800) and the National Natural Science Foundation of China (Nos. 61927806, 62225402, 62321004, 62234001, 12350402, and U22A2074).

## Author contributions

Ning Tang, Junxi Duan and Bo Shen conceived the project. Lei Fu and Yuqing Hu fabricated and transferred the h-BN flakes. Xionghui Jia and Lun Dai performed the AFM measurements. Huaiyuan Yang and Guoping Li discussed the results. Zhuoxian Li, Xiangyan Han and Jianming Lu assisted with the sample fabrication and transfer process. Lei Fu carried out optical measurements and data analysis, and wrote the paper. Ning Tang, Junxi Duan, Weikun Ge and Bo Shen guided the research.

## Competing interests

The authors declare no competing interests.

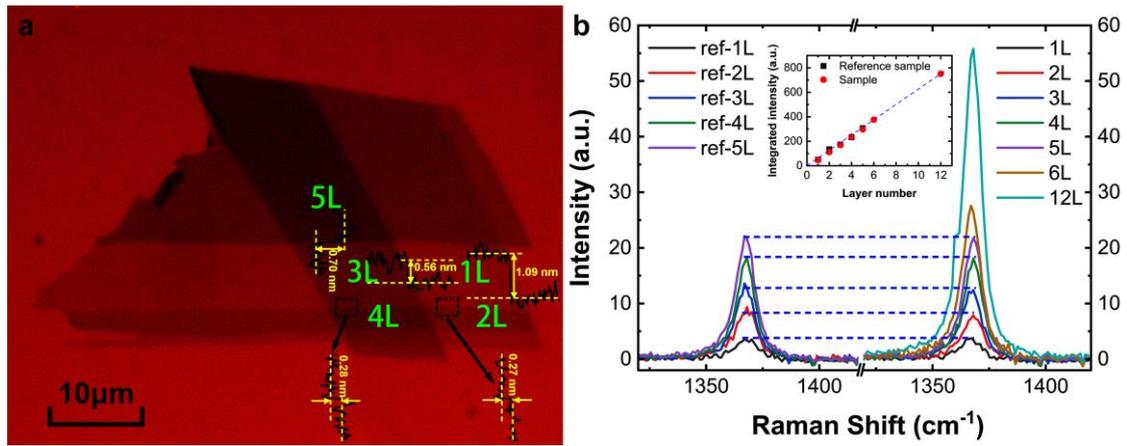

**Fig. 1 Identification of layer numbers of h-BN flakes. a,** Optical image of 1~5L h-BN flake (reference sample) on Si/SiO$_2$ substrate, with contrast adjusted for better visibility, and AFM steps are marked in the image. **b,** The Raman signals of the reference sample (left) and those used for research in this article (right), inset is the integrated intensity of the G band (over 1325 ~ 1400 cm$^{-1}$) with layer numbers, showing apparent linear relationship.

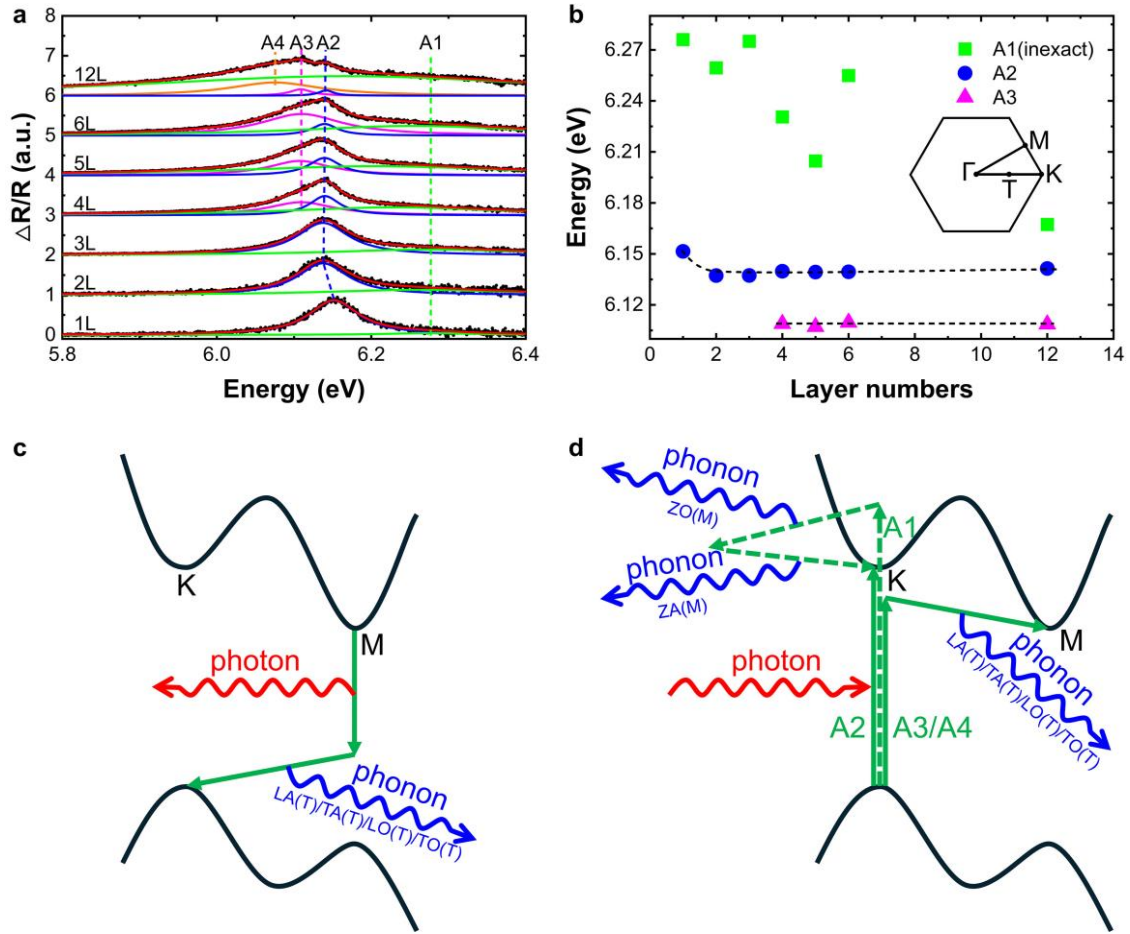

**Fig. 2 Absorption of h-BN with different layer numbers. a,** Reflectance contrast spectrum of various layer numbers of h-BN on sapphire substrate at 4K. The black dots are raw spectrum, the red lines are fitted curves, and other lines are deconvolved peaks, labeled as A1~A4 by the dotted lines, respectively. **b,** The energies of A1~A3 across different layer numbers. Note that A1 peaks are too broad and weak so the fitting energy is inexact. Inset is the first Brillouin zone of h-BN. **c,** Schematic diagram of phonon-assisted recombination processes of the $X_{LA(T)}/X_{TA(T)}/X_{LO(T)}/X_{TO(T)}$ line. **d,** Schematic illustration of direct K-K absorption (A2), phonon-assisted indirect absorption (A3/A4), and phonon-assisted direct absorption (A1).

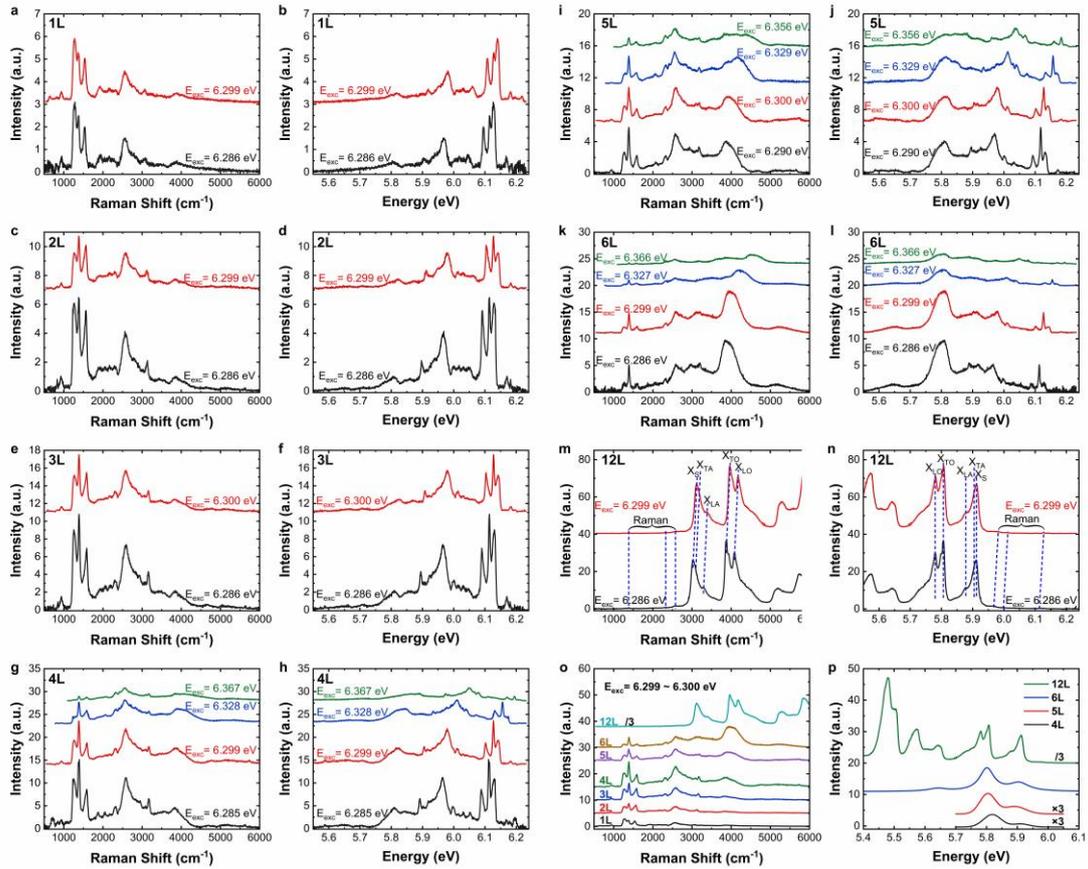

**Fig. 3 Deep-UV photoluminescence and resonance Raman spectroscopies. a-n,** Deep-UV photoluminescence and resonance Raman spectroscopies of 1~6L and 12L under Raman shift coordinates and energy coordinates at 4K. **o,** Comparison of spectrum with different layer numbers under similar excitation Energy of 6.299~6.300eV. **p,** Comparison of the extracted PL components of 4~6L and 12L h-BN.

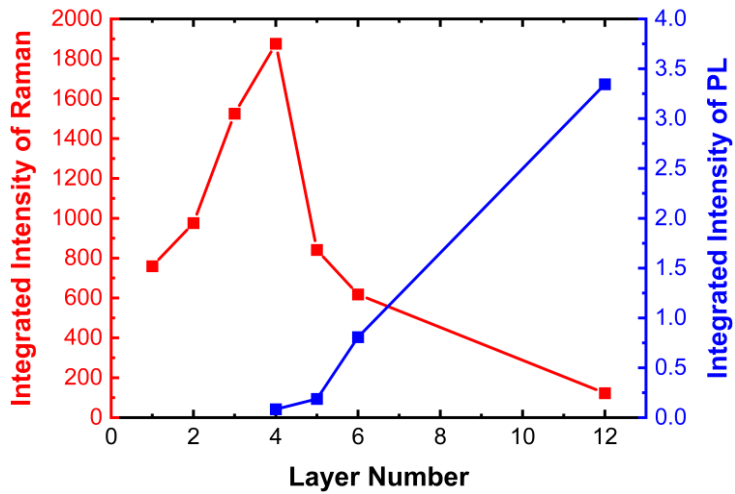

**Fig. 4 The evolution of resonance Raman intensity integrated over 1100~1700 cm$^{-1}$ (red points) and PL intensity integrated over 5.7~6.0 eV (blue points) with different layer numbers.**

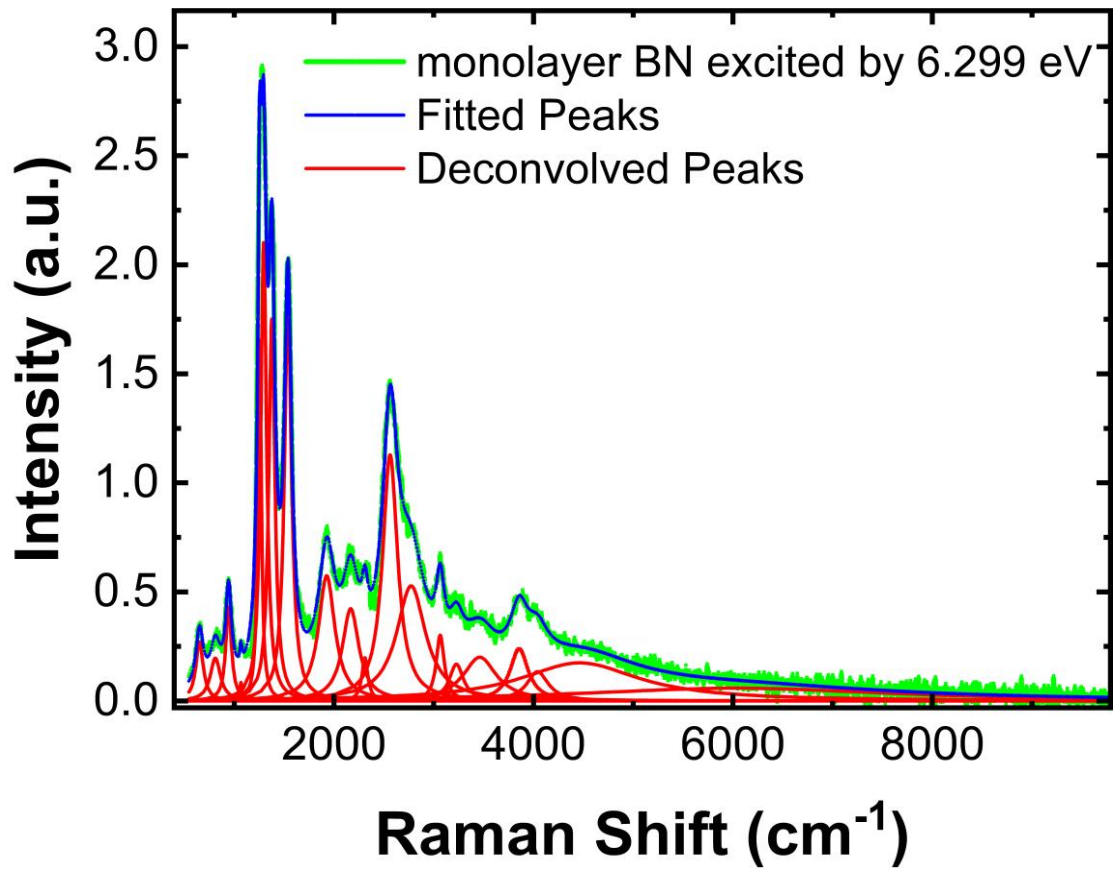

Fig. 5 Deconvolution of resonance Raman spectroscopy of monolayer BN at 4K.

# Unambiguous identification of the indirect band nature of atomically thin hexagonal boron nitride

Lei Fu[1†], Yuqing Hu[4†], Ning Tang[1,2,3*], Junxi Duan[4*], Xionghui Jia[1], Huaiyuan Yang[1], Zhuoxian Li[1], Xiangyan Han[1], Guoping Li[1], Jianming Lu[1], Lun Dai[1,2], Weikun Ge[1], Bo Shen[1,2,3*]

[1]State Key Lab for Mesoscopic Physics and Frontiers Science Center for Nano-optoelectronics, School of Physics, Peking University, Beijing, China.
[2]Collaborative Innovation Center of Quantum Matter, Beijing, China
[3]Peking University Yangtze Delta Institute of Optoelectronics, Nantong, Jiangsu, China
[4]Key Laboratory of Advanced Optoelectronic Quantum Architecture and Measurement (MOE), School of Physics, Beijing Institute of Technology, Beijing, China

[†]Lei Fu and Yuqing Hu contributed equally to this work.
[*]Corresponding author: ntang@pku.edu.cn, junxi.duan@bit.edu.cn, bshen@pku.edu.cn.

**Content**

**Supplementary Note**

**Supplementary Figure**

**Supplementary Note 1. Background subtraction**

To subtract background, we measure the BKG signal just beside the h-BN flakes under the same laser wavelength and power. Then subtract the BKG signal from the acquired sample signal to get the BKG-subtracted signal. Here we take monolayer h-BN under excitation of 6.286eV (197.27nm) for example, as shown in Supplementary Fig. 1. Note that the excitation wavelength of 197.27nm is very close to the cutoff edge of the filter, so there exists a very steep tail. Subtraction of this steep tail results in the poor signal-to-noise (SNR) ratio at Raman shift lower than 1000cm$^{-1}$.

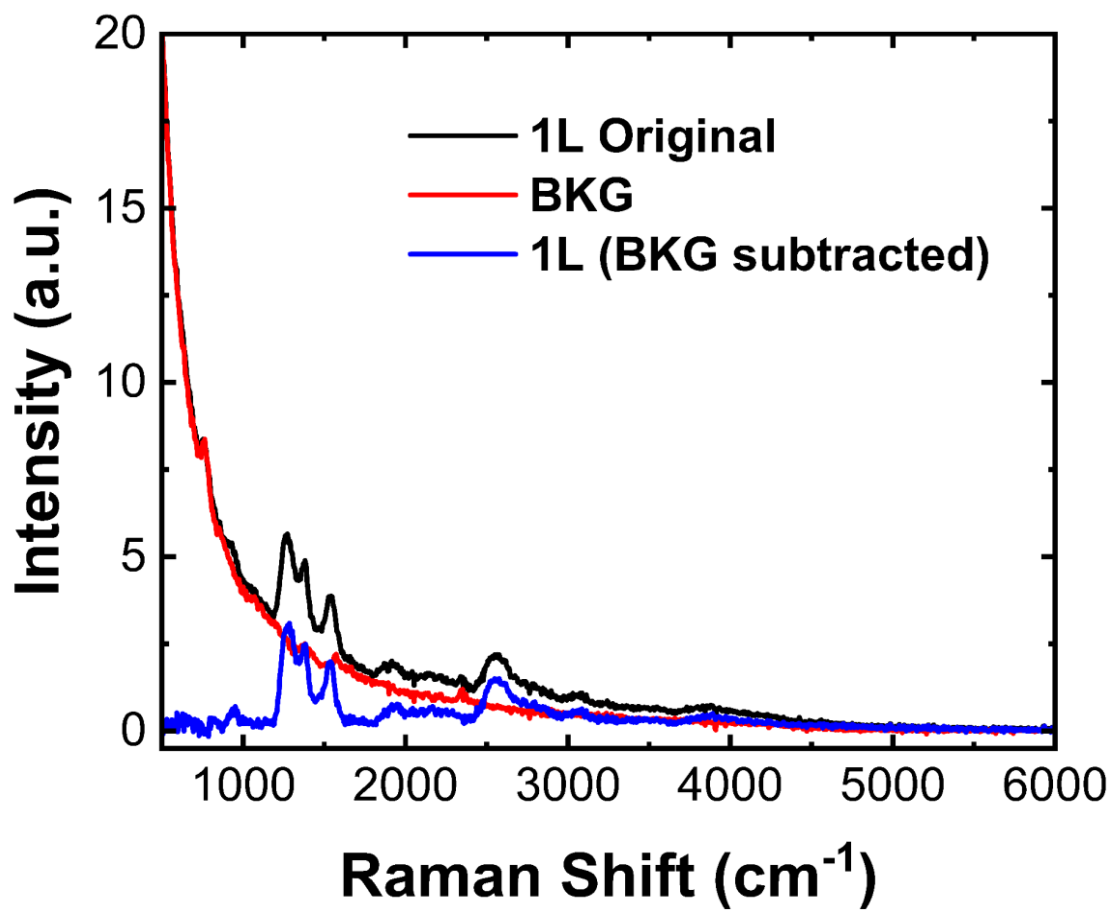

**Supplementary Fig. 1** Background (BKG) subtraction of monolayer h-BN under excitation of 6.286eV (197.27nm).

## Supplementary Note 2. Fine comparison of resonance Raman/PL signals of 1~4L h-BN between different excitation energies

For a better comparison, the resonance Raman/PL signals of 1~4L h-BN under two excitation energies are normalized to the peak at 2562 cm$^{-1}$, as shown in Supplementary Fig. 2. It is clearly seen that the signals match well for 1~3L h-BN, but shows an apparent mismatch around 4000 cm$^{-1}$ for 4L (see the region in the blue circle at Figure S2d), indicating the existence of PL signal.

Note that the intensity between 500~1700cm$^{-1}$ matches not well. This is because this region is close to the main absorption peak around 6.14 eV, and there exist both incident resonance and outgoing resonance for Raman scattering process. A slightly shift of excitation energy can bring about different influence for the outgoing resonance process of Raman peaks in this region.

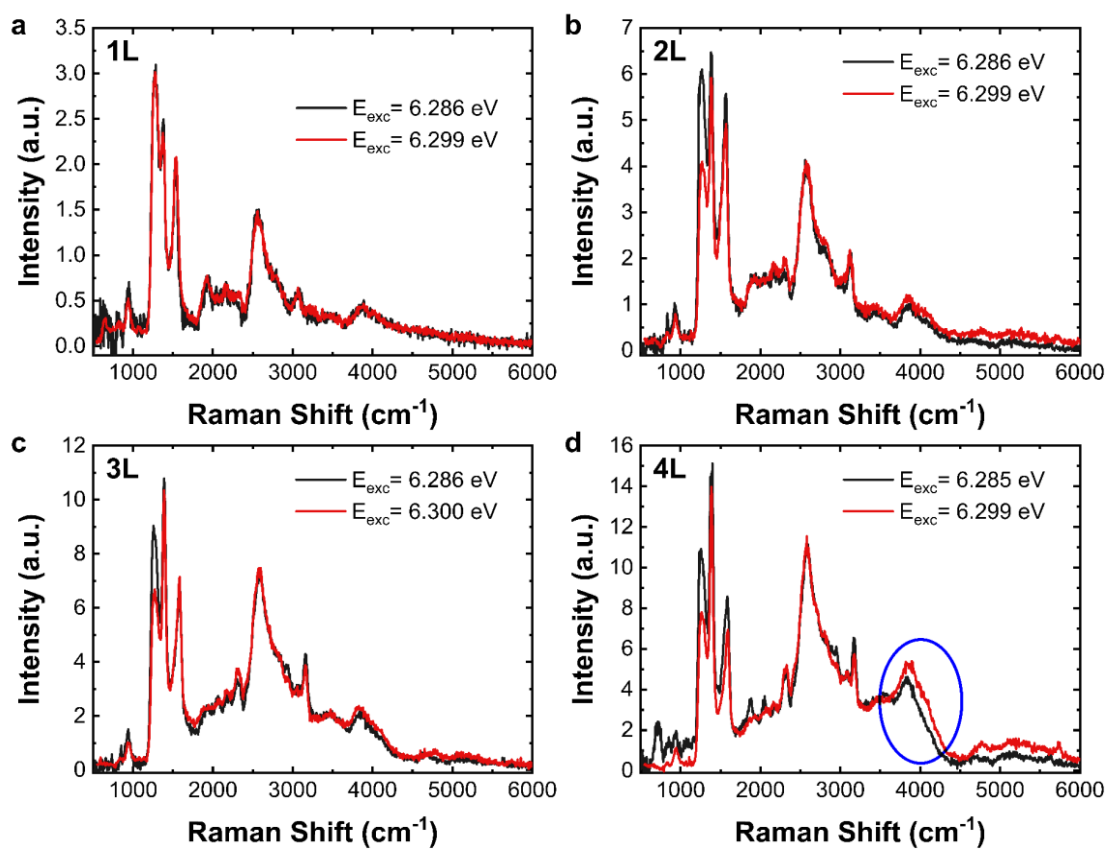

**Supplementary Fig. 2 a-d,** Comparison of resonance Raman/PL signals of 1~4L h-BN upon different excitation energies, by normalizing the spectrum to the 2562 cm$^{-1}$ peak.

## Supplementary Note 3. Fine comparison of resonance Raman/PL signals between 1~4L h-BN under excitation energy of 6.299~6.300eV

The resonance Raman/PL spectrum of 1~6L h-BN under the same excitation energy are normalized to the peak at 2562 cm$^{-1}$, as shown in Supplementary Fig. 3. The spectrum in the 1800~6000cm$^{-1}$ range match well for 1~3L h-BN, neglecting some minor differences, whereas that of 4~6L h-BN shows an apparent mismatch around 4000 cm$^{-1}$, thus confirming the rationality of subtract the spectrum of 3L from the spectrum of 4~6L, to obtain the approximate PL spectrum of 4~6L.

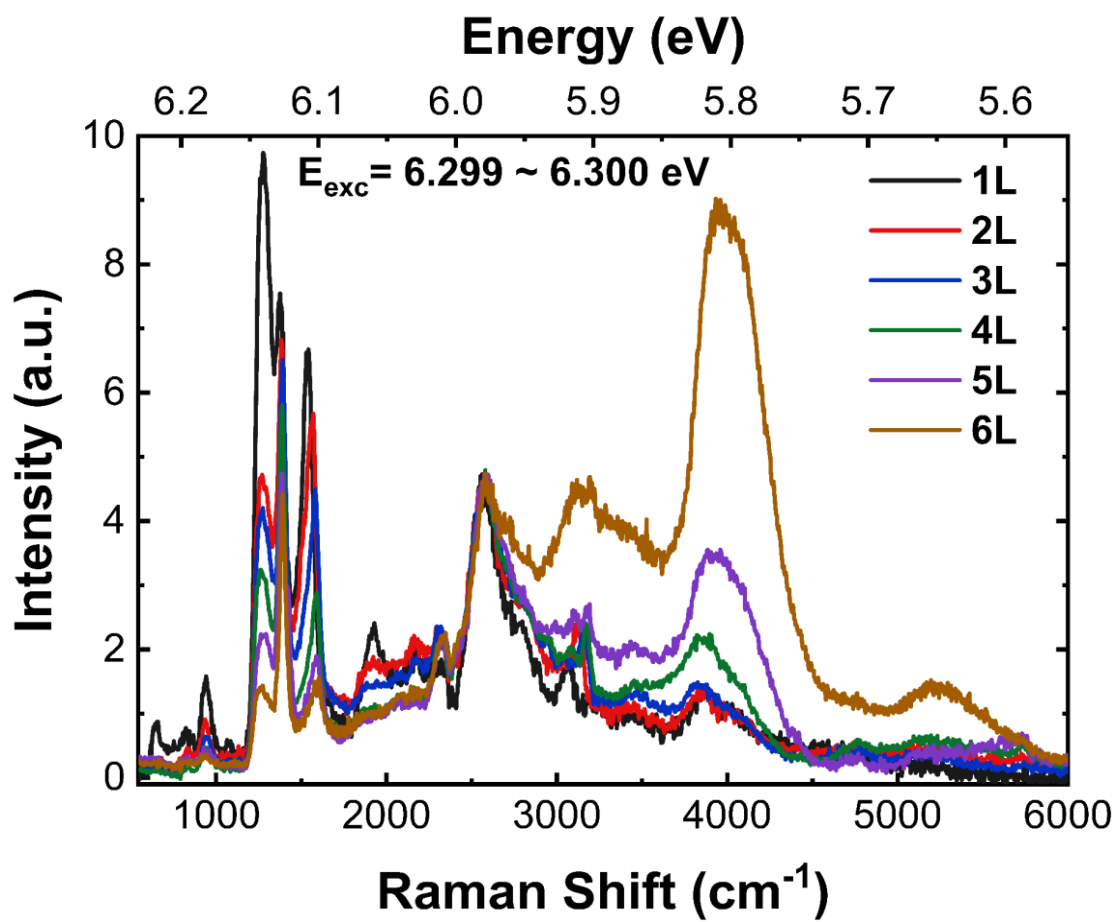

**Supplementary Fig. 3** Comparison of the resonance Raman/PL signals between 1~6L h-BN under the same excitation energy of 6.299~6.300 eV, by normalizing the spectrum to the 2562 cm$^{-1}$ peak.

**Supplementary Note 4. Extraction of PL components for 4~6L h-BN**

We roughly subtract the spectrum of 3L (normalized to the 2562 cm$^{-1}$ peak of 4~6L) from the spectrum of 4~6L, followed by removing spikes, and fitting by Gaussian function (for 4~5L) and Lorentz function (for 6L), respectively. Finally, the PL components of 4~6L are roughly extracted, as shown in Supplementary Fig. 4.

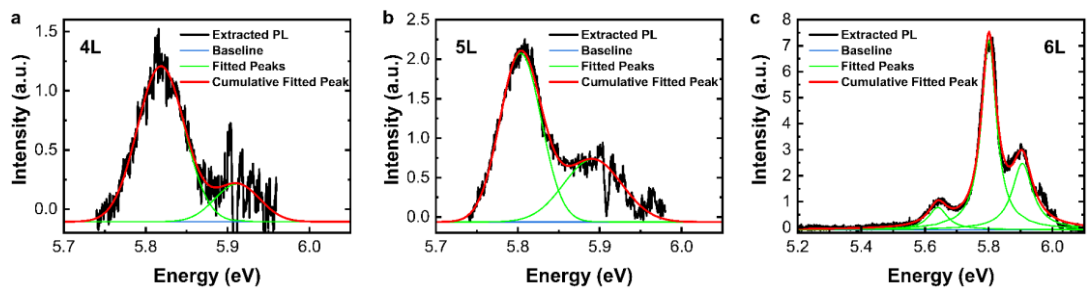

**Supplementary Fig. 4** Fitting process of the PL components of 4~6L h-BN (a-c). The black lines are raw PL signals subtracted by the spectrum of 3L, with some spikes removed. The green lines are fitted peaks, and the red lines are cumulative fitted peaks.